# Effects of Important Parameters Variations on Computing Eigenspace-Based Minimum Variance Weights for Ultrasound Tissue Harmonic Imaging

Mehdi Haji Heidari [a], Moein Mozaffarzadeh[a,b,*], Rayyan Manwar[c], and Mohammadreza Nasiriavanaki[c]

[a]Department of Biomedical Engineering, Tarbiat Modares University, Tehran, Iran
[b]Research Center for Biomedical Technologies and Robotics, Institute for Advanced Medical Technologies, Tehran, Iran
[c]Department of Biomedical Engineering, Wayne State University, Detroit, Michigan, USA

**ABSTRACT**

In recent years, the minimum variance (MV) beamforming has been widely studied due to its high resolution and contrast in B-mode Ultrasound imaging (USI). However, the performance of the MV beamformer is degraded at the presence of noise, as a result of the inaccurate covariance matrix estimation which leads to a low quality image. Second harmonic imaging (SHI) provides many advantages over the conventional pulse-echo USI, such as enhanced axial and lateral resolutions. However, the low signal-to-noise ratio (SNR) is a major problem in SHI. In this paper, Eigenspace-based minimum variance (EIBMV) beamformer has been employed for second harmonic USI. The Tissue Harmonic Imaging (THI) is achieved by Pulse Inversion (PI) technique. Using the EIBMV weights, instead of the MV ones, would lead to reduced sidelobes and improved contrast, without compromising the high resolution of the MV beamformer (even at the presence of a strong noise). In addition, we have investigated the effects of variations of the important parameters in computing EIBMV weights, i.e., $K$, $L$, and $\delta$, on the resolution and contrast obtained in SHI. The results are evaluated using numerical data (using point target and cyst phantoms), and the proper parameters of EIBMV are indicated for THI.

## 1. INTRODUCTION

Second Harmonic Imaging (SHI) is a well-known ultrasound medical imaging technique that has improved image quality in many clinical ultrasound applications.[1] The resolution and contrast provided by the harmonic imaging is much better than the fundamental imaging.[2,3] Although the harmonic imaging provides advantages compared to the fundamental imaging, the amplitudes of the harmonic components are so weak. Therefore, Tissue Harmonic Imaging (THI) always suffers from a poor Signal-to-Noise Ratio (SNR).[4–6]

Beamforming plays a significant role in the quality of the formed images. There are a large number of publications in this field of study.[7–9] As novel weighting methods, two modifications of the Coherence Factor (CF) have been introduced.[10,11] The Minimum Variance (MV) beamformer is an adaptive weighting method which can be applied to the medical ultrasound B-mode images to improve the resolution.[12] The MV relies on weighted summation of received signals in which weights are calculated based on the information obtained by the spatial properties of the recorded signals. There are a number of modifications for MV algorithm.[13–18] Adaptive beamforming methods used in ultrasound images provides a significant higher resolution. However, the contrast enhancement is not investigated significantly. The Eigenspace-Based Minimum Variance (EIBMV) beamformer is a technique in which not only it can improve the resolution, but also significantly enhances the contrast, compared to Delay and Sum (DAS) and MV beamformers. The weight vector of the EIBMV is generated by projecting the MV weight vector onto a vector subspace constructed from the eigenstructure of the covariance

---

matrix.[19] In the EIBMV approach, the effective parameters for computing weights have standard values which are generally applied to the MV-based approaches.[19] In this paper, the EIBMV method has been applied to SHI in order to overcome the powerful noise, resulting in contrast improvement. In addition, we have investigated the effects of variations of the important parameters for computing EIBMV weights on the resolution and contrast in THI, and finally, the appropriate parameters are presented.

The rest of the paper is organized as follows. The structure of the EIBMV beamformer and the procedure of choosing the appropriate weights vector are explained in the section 2. The influence of the variations of EIBMV parameters along with the optimal values for parameters and the corresponding results are illustrated in the section 3. Finally, the discussion and conclusion are presented in the section 4.

## 2. METHODS

For a linear array with $M$ elements, the output of a general beamformer is defined as follow:

$$z(k) = \mathbf{w}^H(k)\mathbf{y}(k) = \sum_{i=1}^{M} w_i * (k) y_i(k - \Delta_i), \qquad (1)$$

where $y(k) = [y_1(k - \Delta_1), ..., y_M(k - \Delta_M)]$ is the time delayed version of array observation corresponding to a specific point of the image, $\Delta_i$ is the time-delay applied to channel $i$, $\mathbf{w} = [w_1, ..., w_M]^T$ is the complex vector of beamformer weight, and the superscripts $(.)^*$ and $(.)^H$ denote the conjugate and conjugate transpose, respectively. In the DAS beamformer, the weights are predetermined by some windowing functions such as rectangular, Hanning and Hamming windows. In the MV beamformer, the optimum weight vectors are obtained as a function of the spatial covariance matrix, $\mathbf{R}$, as follows:

$$\mathbf{w}_{MV} = \frac{\mathbf{R}^{-1}\mathbf{a}}{\mathbf{a}^H \mathbf{R}^{-1} \mathbf{a}}, \qquad (2)$$

where $\mathbf{R}(k) = E\{\mathbf{y}(k)\mathbf{y}(k)^H\}$ is the $M \times M$ spatial covariance matrix, and $\mathbf{a}$ is a vector of all ones with length in accordance with $\mathbf{R}$. In practice, the exact covariance matrix, $\mathbf{R}$, is unavailable. Hence, the covariance matrix at time $k$ is estimated by using temporal averaging and spatial smoothing techniques, where spatial smoothing is done through division of the array into subarrays to obtain a good estimate of the covariance matrix.[18] $\mathbf{R}$ can be estimated as follows:

$$\hat{\mathbf{R}}(k) = \frac{1}{(2K+1)(M-L+1)} \sum_{n=-K}^{K} \sum_{i=1}^{M-L+1} \mathbf{y}^i(k+n)\mathbf{y}^i(k+n)^H, \qquad (3)$$

where $L$ is the subarray length, $\mathbf{y}^i(k+j)$ is the delayed input vector for the $i^{th}$ subarray, and $K$ is the temporal averaging number. To insure a well-conditioned covariance matrix and improve the beamformer robustness, the covariance matrix is diagonally loaded, $\hat{\mathbf{R}}(k) = \hat{\mathbf{R}}(k) + \varepsilon \mathbf{I}$, where amount of the $\varepsilon$ has been set $\Delta$ times of the power in the received signals according to $\varepsilon = \Delta \text{trace}\{\hat{\mathbf{R}}(k)\}$, and $\mathbf{I}$ is the identify matrix.[12,20] The MV beamformer is commonly used in ultrasound imaging in order to improve resolution of the images. However, the contrast enhancement has no salient progress. EIBMV method, using the eigenstructure of the covariance matrix besides preserving the desire signal, suppresses the sidelobes and consequently improves image quality in the terms of contrast and resolution. Hence, EIBMV approach retains the resolution while improves the contrast, compared to MV and DAS beamformers. In this technique, the estimated covariance matrix is broken to two orthogonal spaces i.e., signal subspace and noise plus interference one. The diagonally loaded covariance matrix $\mathbf{R}$ can be written in terms of its Eigen decomposition as:

$$\hat{\mathbf{R}} = \mathbf{V} \mathbf{\Lambda} \mathbf{V}^H, \qquad (4)$$

where $\mathbf{\Lambda} = diag[\lambda_1, \lambda_2, ..., \lambda_L]$ so that $\lambda_1 \geq \lambda_2 \geq ... \geq \lambda_L$ and $\mathbf{V} = [\mathbf{v}_1, \mathbf{v}_2, ..., \mathbf{v}_L]$ in which $\mathbf{v_i}, i = 1, 2, ..., L$ are the orthonormal eigenvectors corresponding to $\lambda_i, i = 1, 2, ..., L$. To limit the effects of interferences and noise, the covariance matrix can be generated by a few numbers of eigenvectors. The signal subspace $\mathbf{E}_s$ only contains the desire signal and significantly decreases the sidelobes effects. This subspace is constructed from the eigenvectors corresponding to the largest eigenvalues.

Table 1: Parameters of the Simulation Setup.

| Value | Parameter |
|---|---|
| 50 $MHz$ | Sampling frequency ($f_s$) |
| 1.96 $MHz$ | Center fundamental frequency ($f_0$) |
| 3.92 $MHz$ | Center fundamental frequency ($2f_0$) |
| 132 | Number of transmission elements |
| 66 | Number of receive elements ($M$) |
| 409 $\mu m$ | Pitch |
| 20 $\mu m$ | Kerf |
| 1540 $m/s$ | speed of sound ($C_0$) |
| 50 $mm$ | Focal depth (transmit focus) |
| 1000 $kPa$ | Initial pressure |
| 1000 $kg/m^3$ | Medium density |
| 0.5 $dB/cm/MHz$ | attenuation coefficient |
| 3.5 | Nonlinear parameter ($\beta$) |
| 2 | Gamma coefficient ($\gamma$) |

Table 2: Parameters for Weights Calculation.

| Parameter | First | Second | Third | Fourth | Fifth |
|---|---|---|---|---|---|
| $K$ | 0 | $\frac{1}{3}K_{stan}$ | $\frac{1}{2}K_{stan}$ | $K_{stan}$ | $2K_{stan}$ |
| $L$ | 1 | $\frac{1}{6}M$ | $\frac{1}{3}M$ | $\frac{1}{2}M = L_{stan}$ | $M$ |
| $\Delta$ | 0 | $\frac{1}{100}\Delta_{stan}$ | $\frac{1}{10}\Delta_{stan}$ | $\Delta_{stan}$ | $10\Delta_{stan}$ |
| $\delta$ | 0 | 0.1 | 0.5 | 0.8 | 1 |

$$\mathbf{E}_s = [\mathbf{v}_1, ..., \mathbf{v}_{Num}] \quad (5)$$

where $Num$ is the number of the eigenvectors effectively demonstrate the signal subspace. One metric for determining the $NUM$ could be defined as the number of dominant eigenvalues, the amount of which are larger than $\delta\lambda_{\max}$, where $\delta$ is a positive factor smaller than 1, and $\lambda_{\max}$ is the largest eigenvalue. Finally, the weight vector of the EIBMV is given by projecting the MV weight vector onto signal subspace.[19]

$$\mathbf{w}_{EIBMV} = \mathbf{E}_s\mathbf{E}_s^H\mathbf{w}_{MV} \quad (6)$$

## 3. RESULTS

### 3.1 Data Acquisition

In both MV and EIBMV adaptive beamformers, the effective parameters for weights calculation have standard values; these parameters are: (1) the number of samples for the temporal averaging of the covariance matrix ($K$), (2) the number of subarray elements for computing the covariance matrix ($L$), (3) the diagonal loading factor ($\Delta$), and (4) In EIBMV approach the $\delta$ parameter, which their standard values are the length of the transmission pulse ($K_{stan}$), half of the number of array elements ($L = M/2$), ($K_{stan} = 1/100L$), and $\delta_{stan} = 0.5$, respectively.[19] In this study, we aim to investigate the influence of all the four effective parameters of EIBMV weights calculation on the resolution and contrast, for the SHI, in order to choose their best value. In this section, we have simulated two phantoms, a phantom of wire targets located at the depth of 22.5-62.5 $mm$, separated by 5 $mm$ at each depth, and a phantom including 5 cysts with an equal diameter of 5 $mm$, located at the depth of 22.5-62.5 $mm$, separated by 10 $mm$. The phantoms are used in order to investigate the performance of the EIBMV beamformer in the terms of the lateral resolution and contrast, along with changes of $K$, $L$, $\Delta$ and $\sigma$. The simulations have been carried out using CREANUIS, which is a non-linear radio frequency (RF) ultrasound image simulator.[21,22] In the all simulations, a 132-element curvilinear array transducer were used. In transmission, all elements are participated while in reception only the nearest 66 elements of the transducer which were placed in front of the scan line were used. A two-cycle Gaussian weighted Sinusoidal signal was used in transmission, and the transmit focus was set at 50 $mm$ depth. Other parameters for simulations are listed in Table 1. In all the simulations, to separate the harmonic from fundamental frequencies, we have used PI technique which is a well-known technique.[23] Before separating the harmonic from fundamental by the PI method, white Gaussian noise with a SNR of 60 $dB$ was added to the received signal. In the MV and EIBMV beamformers, the amounts of parameters considered for computing weights are listed at Table 2.

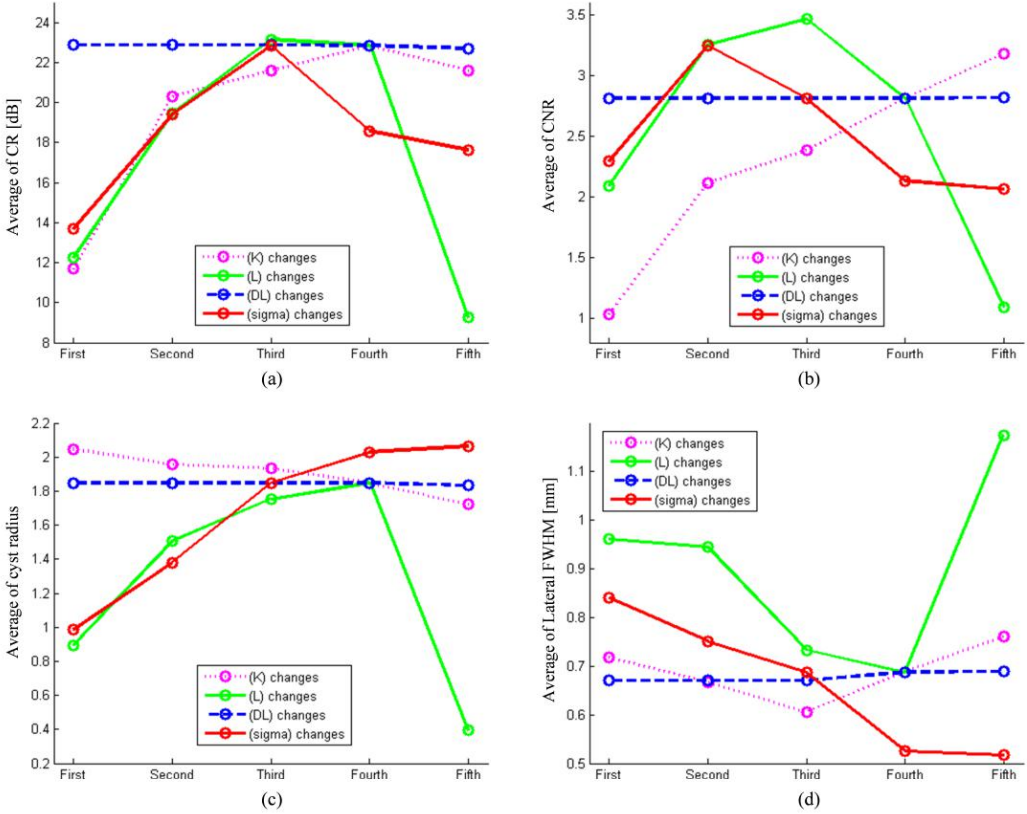

Figure 1: The patterns obtained from the average of (a) CR, (b) CNR, and (c) estimated cyst radius (for the simulated cyst phantom along the depths of 22.5-62.5 $mm$), and (d) the average of lateral FWHM (for 9 points along the depths of 22.5-62.5 $mm$) for variations of the $K$, $L$, $\delta$, and $\Delta$ parameters in the weight vector construction, using a transmit focus of 50 $mm$, 1.96 $MHz$ curvilinear array. The amounts of $K$, $L$, $\delta$, and $\Delta$ parameter considered for computing weights are listed at Table 2.

### 3.2 Simulated Point Targets

Here, we aim to explore the influence of variations of the effective parameters of the EIBMV on the resolution of the output image. We have used the Full Width at Half Maximum (FWHM) index to evaluate the mainlobe width of the images for each point. Fig. 1(d) shows the influence of variations of $K$, $L$, $\Delta$ and $\delta$ on the average of the mainlobe width (in the term of FWHM), for all the target points. It is notable that a less mainlobe width for a point in a specific depth indicates a higher resolution.

### 3.3 Simulated Cyst Targets

To investigate the influence of the variations of the effective parameters in EIBMV, in the term of contrast, a phantom of cysts were used. In simulating the cyst phantom, the speckle distribution is simulated randomly with

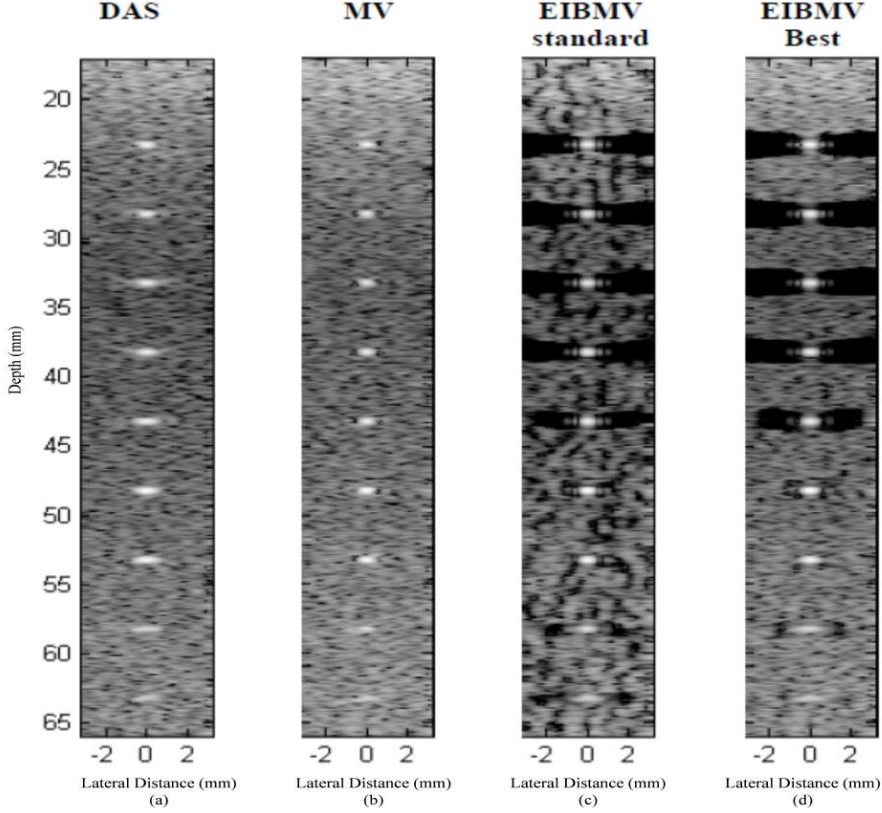

Figure 2: Simulated point targets using a transmit focus 50 $mm$, 1.96 $MHz$ curvilinear array. (a) DAS, (b) MV, (c) EIBMV with standard parameters, and (d) EIBMV with the best parameters. All images are shown with a dynamic range of 50 $dB$.

a density of 10 scatters per a volume $\lambda^3$, where $\lambda$ is the wavelength of the central frequency of the transmitted pulse, to ensure the speckles are fully developed.[24] The scattering amplitudes are randomly distributed between 0 and 1. To evaluate the contrast parameters of the methods, we have used the contrast ratio (CR) and the contrast-to-noise ratio (CNR) metrics. The CR is defined as the difference between the mean value in the background and the mean value in the cyst region in dB,[7] and the CNR is a measurement defined as the ratio of the CR to the standard deviation of the image intensity in the background region.[20] To compute the mean values and the standard deviation of the image intensities for each cyst, we considered two circles with exact radius of cyst, located at the same depth of the cyst and at the lateral positions, -3 $mm$ and +3 $mm$, to determine the region of the cyst and its background, respectively. Moreover, to verify the edge definition, we used indices, the mean of radius estimation, too. The influence of variations of $K$, $L$, $\delta$, and $\Delta$ parameters in constructing weight vector of EIBMV on the average of variations CR, CNR and mean of radius estimation are shown in Fig. 1(a), Fig. 1(b) and Fig. 1(c), respectively. The results presented in Fig. 1 show that there are a heavy trade off between resolution, CR, CNR, and the radius estimation in choosing the best value of EIBMV beamformer parameters. Hence, we have to consider all aspects, and by prioritizing them, choose the best value of the EIBMV beamformer parameters. As mentioned earlier, although the resolution of harmonic imaging is satisfied, the contrast is not well due to the weakness of the amplitudes of the harmonic components and SNR. Therefore, it is better to give the highest priority to contrast parameters (CR and CNR). The amount of $K$ is not very effective in the resolution and cyst radius while in the case of CR and CNR, the fourth and fifth cases have good results. Thus, we can consider the average of the amounts of related to fourth and fifths cases as the optimum value for $K$. The amount of $L$ in all the terms of resolution, CR, CNR, and cyst radius has the optimal common values (the fourth and fifth). Due to considering the higher importance of CNR compared to the cyst radius, we choose the amount the third case as the optimum value for $L$. The optimal common values of $\sigma$, in cases of resolution and

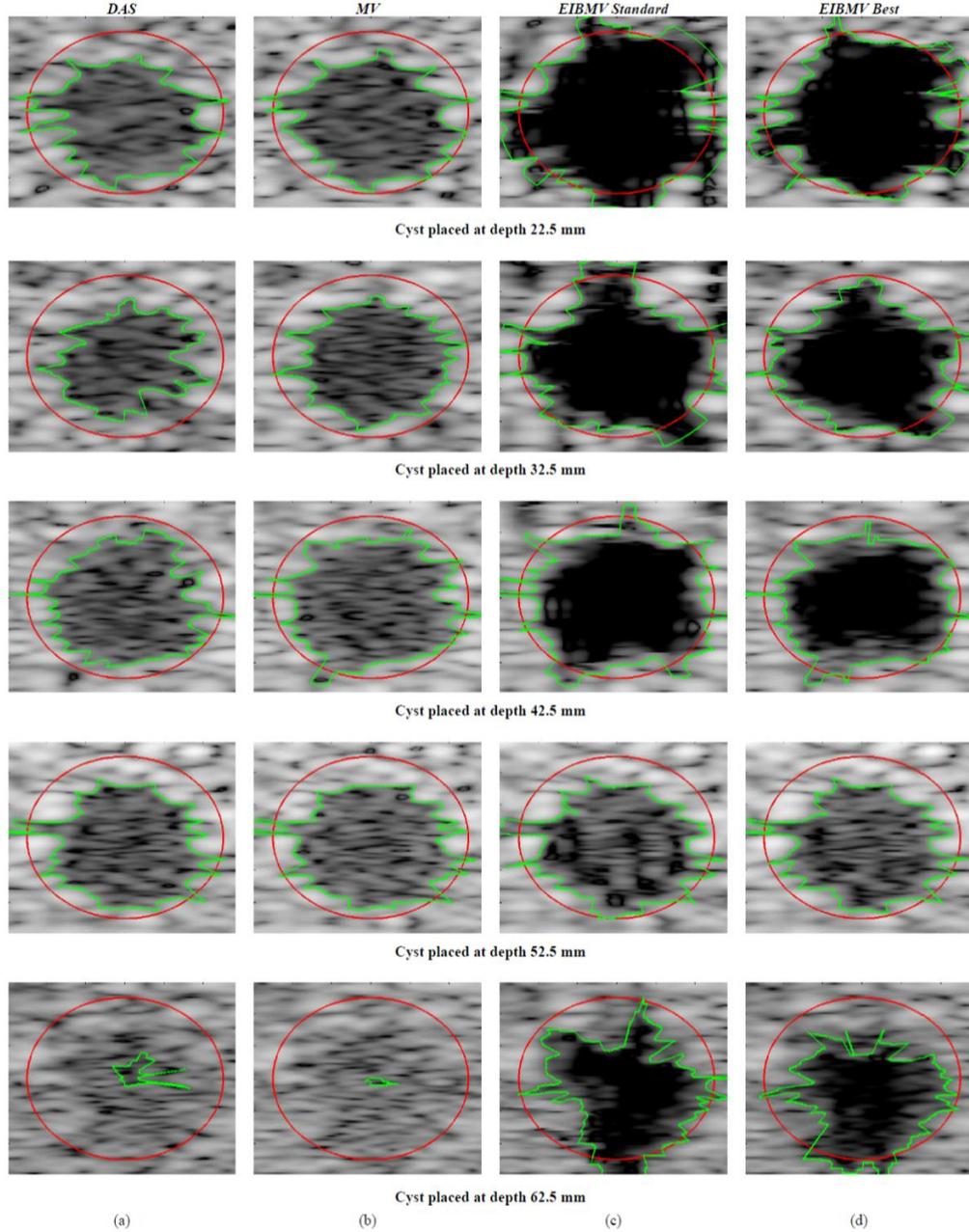

Figure 3: Boundaries of simulated cyst phantom using a transmit focus 50 $mm$, 1.96 $MHz$ curvilinear array. (a) DAS, (b) MV, (c) EIBMV with standard parameters, and (d) EIBMV with best parameters. The panels from top to bottom are cyst located at depths 22.5 $mm$, 32.5 $mm$, 42.5 $mm$, 52.5 $mm$, and 62.5 $mm$, respectively. All the images are shown with a dynamic range of 50 $dB$.

cyst radius are the values of the fourth and fifth cases, while for CR and CNR, the values are selected based on the second and third cases. Due to considering a higher importance for CR and CNR, compared to resolution and cyst radius parameter, we choose the amount related to third case as the optimum value for $\delta$, which is the optimal common value among all the resolution and contrast parameters. The $\Delta$ parameter does not have significant effect on the resolution and contrast parameters. So, we totally ignore analysis of this parameter; and set its optimal value as its standard value. Results demonstrate that the best values of parameters $K$, $L$, $\Delta$ and $\delta$ for the EIBMV beamformer are $K_{best} = 1.5 K_{stan}$, $L_{best} = 2/3 L_{stan} = 1/3 M$, $\Delta_{best} = \Delta_{stan}$ and $\delta_{best} = \delta_{stan}$,

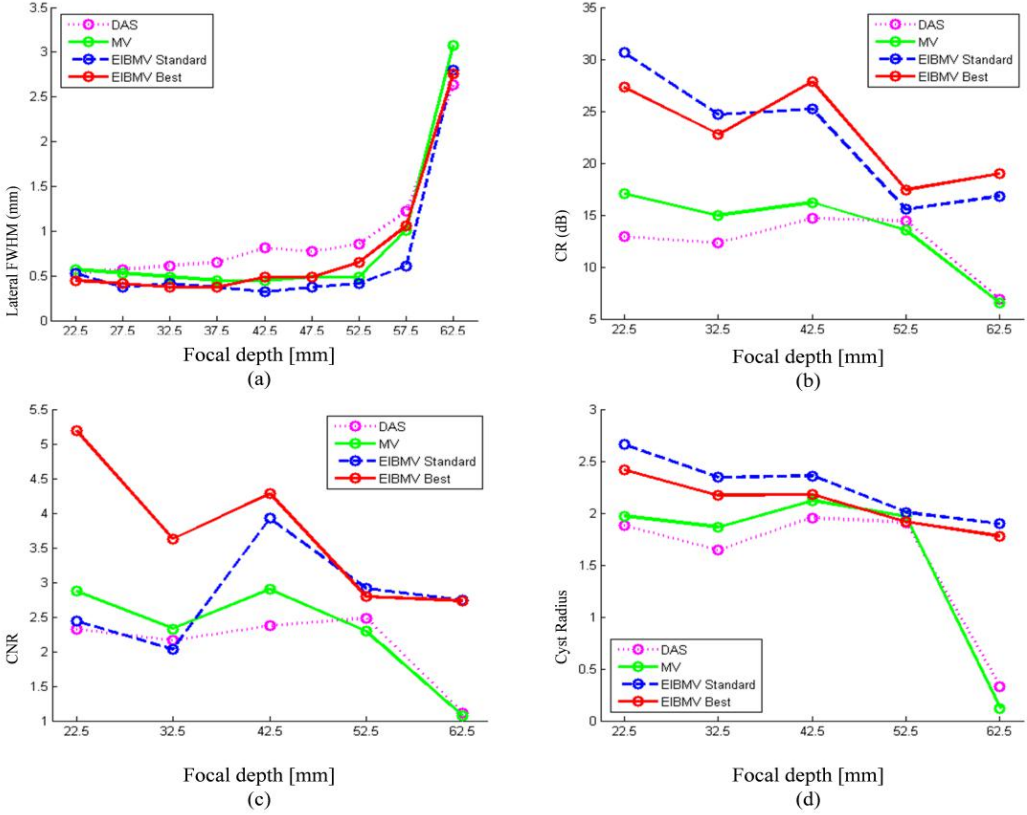

Figure 4: The patterns of (a) Lateral FWHM for 9 points along depths 22.5-62.5 $mm$, and the patterns of (b) CR, (c) CNR, and (d) estimated cyst radius, from simulated cyst phantom along depths 22.5-62.5 $mm$, respectively, for (a) DAS, (b) MV, (c) EIBMV with standard parameters, and (d) EIBMV with best parameters, using a transmit focus 50 $mm$, 1.96 $MHz$ curvilinear array.

respectively.

## 4. DISCUSSION AND CONCLUSION

Harmonic imaging has good axial and lateral resolutions. However, the low SNR is a major problem involved with this imaging method. In this paper, EIBMV beamformer is applied to SHI. Using EIBMV weights instead of the MV ones, leads to reduced sidelobes and improved contrast, without compromising the high resolution of the MV beamformer, even at the presence of strong noise. Moreover, here we investigated the effects of variations of the important parameters in computing EIBMV weights, i.e., $K$, $L$, $\Delta$ and $\delta$, on the resolution and contrast while THI is performed. To this end, the effects of each parameter is separately explored while other parameters were fixed (having their standard values). The results, using point target and cyst phantom, are evaluated on numerical data, and the proper parameters of EIBMV are indicated for THI. Results demonstrate that the best values of $K$, $L$, $\Delta$ and $\delta$ parameters for EIBMV beamformer are $K_{best} = 1.5 K_{stan}$, $L_{best} = 2/3 L_{stan} = 1/3M$, $\Delta_{best} = \Delta_{stan}$ and $\delta_{best} = \delta_{stan}$, respectively. Fig. 2 and Fig. 3 show the images obtained from the mentioned phantoms in the section 2, point targets and cyst phantom, respectively, using different beamformers (DAS, MV, EIBMV with standard parameters, and EIBMV with the best parameters), displayed over a 50 $dB$ dynamic range. In Fig. 3 the boundaries of the simulated cyst phantoms are displayed, at all depth, over a 30 $dB$ dynamic range. The red circles and the green curves plotted in Figs. 3 represent the true and estimated border of the cysts, respectively. The FWHM variation along the depths for all points of Fig. 2, and also the variations of relative CR, CNR, mean of radius estimation, along the depths for all cyst targets of Fig. 3, are shown in Fig. 4, respectively. As the results show, although EIBMV method, where the proposed parameter are used instead of the standard parameter, decreases the ability to improve the resolution and cyst radius, it improves CR and CNR significantly. In other words, the EIBMV beamformer could obtain a better contrast by sacrificing the resolution, which would be suitable for harmonic imaging. The presented results show that, in average, EIBMV method when the proposed parameter are used provides CR enhancement of 10.6 $dB$, 9.2 $dB$, and 0.4 $dB$, and CNR improvement of about 78%, 62%, and 32% in comparison with DAS, MV and EIBMV with standard parameter, respectively. It would be more efficient to implement the EIBMV technique with the MV methods having a low computational complexities[25–29] instead of using the conventional MV. Therefore, a practical method can be achieved. This is a topic for our future work.

# REFERENCES

[1] Li, Y. and Zagzebski, J. A., "Computer model for harmonic ultrasound imaging," *IEEE transactions on ultrasonics, ferroelectrics, and frequency control* **47**(4), 1000–1013 (2000).

[2] Shen, C.-C. and Li, P.-C., "Harmonic leakage and image quality degradation in tissue harmonic imaging," *IEEE transactions on ultrasonics, ferroelectrics, and frequency control* **48**(3), 728–736 (2001).

[3] Bouakaz, A. and de Jong, N., "Native tissue imaging at superharmonic frequencies," *IEEE transactions on ultrasonics, ferroelectrics, and frequency control* **50**(5), 496–506 (2003).

[4] Du, Y., Rasmussen, J., Jensen, H., and Jensen, J. A., "Second harmonic imaging using synthetic aperture sequential beamforming," in [*Ultrasonics Symposium (IUS), 2011 IEEE International*], 2261–2264, IEEE (2011).

[5] Christopher, T., "Finite amplitude distortion-based inhomogeneous pulse echo ultrasonic imaging," *IEEE transactions on ultrasonics, ferroelectrics, and frequency control* **44**(1), 125–139 (1997).

[6] Mardi, Z. and Mahloojifar, A., "Tissue second harmonic ultrasound imaging using huffman sequence," in [*Biomedical Engineering and 2016 1st International Iranian Conference on Biomedical Engineering (ICBME), 2016 23rd Iranian Conference on*], 182–186, IEEE (2016).

[7] Mozaffarzadeh, M., Mahloojifar, A., Orooji, M., Adabi, S., and Nasiriavanaki, M., "Double-stage delay multiply and sum beamforming algorithm: Application to linear-array photoacoustic imaging," *IEEE Transactions on Biomedical Engineering* **65**(1), 31–42 (2018).

[8] Mozaffarzadeh, M., Mahloojifar, A., and Orooji, M., "Image enhancement and noise reduction using modified delay-multiply-and-sum beamformer: Application to medical photoacoustic imaging," in [*Iranian Conference on Electrical Engineering (ICEE) 2017*], 65–69, IEEE (2017).

[9] Mozaffarzadeh, M., Sadeghi, M., Mahloojifar, A., and Orooji, M., "Double-stage delay multiply and sum beamforming algorithm applied to ultrasound medical imaging," *Ultrasound in Medicine and Biology* **44**(3), 677 – 686 (2018).

[10] Mozaffarzadeh, M., Yan, Y., Mehrmohammadi, M., and Makkiabadi, B., "Enhanced linear-array photoacoustic beamforming using modified coherence factor," *Journal of Biomedical Optics* **23**(2), 026005 (2018).

[11] Mozaffarzadeh, M., Mehrmohammadi, M., and Makkiabadi, B., "Image improvement in linear-array photoacoustic imaging using high resolution coherence factor weighting technique," *arXiv preprint arXiv:1710.02751* (2017).

[12] Näsholm, S. P., Austeng, A., Jensen, A. C., Nilsen, C.-I. C., and Holm, S., "Capon beamforming applied to second-harmonic ultrasound experimental data," in [*Ultrasonics Symposium (IUS), 2011 IEEE International*], 2217–2220, IEEE (2011).

[13] Mozaffarzadeh, M., Mahloojifar, A., and Orooji, M., "Medical photoacoustic beamforming using minimum variance-based delay multiply and sum," in [*Digital Optical Technologies 2017*], **10335**, 1033522, International Society for Optics and Photonics (2017).

[14] Mozaffarzadeh, M., Avanji, S. A. O. I., Mahloojifar, A., and Orooji, M., "Photoacoustic imaging using combination of eigenspace-based minimum variance and delay-multiply-and-sum beamformers: Simulation study," *arXiv preprint arXiv:1709.06523* (2017).

[15] Paridar, R., Mozaffarzadeh, M., Mehrmohammadi, M., and Orooji, M., "Photoacoustic image formation based on sparse regularization of minimum variance beamformer," *arXiv preprint arXiv:1802.03724* (2018).

[16] Paridar, R., Mozaffarzadeh, M., Nasiriavanaki, M., and Orooji, M., "Double minimum variance beamforming method to enhance photoacoustic imaging," *arXiv preprint arXiv:1802.03720* (2018).

[17] Mozaffarzadeh, M., Mahloojifar, A., Orooji, M., Kratkiewicz, K., Adabi, S., and Nasiriavanaki, M., "Linear-array photoacoustic imaging using minimum variance-based delay multiply and sum adaptive beamforming algorithm," *Journal of Biomedical Optics* **23**(2), 026002 (2018).

[18] Mozaffarzadeh, M., Mahloojifar, A., Nasiriavanaki, M., and Orooji, M., "Eigenspace-based minimum variance adaptive beamformer combined with delay multiply and sum: Experimental study," *arXiv preprint arXiv:1710.01767* (2017).

[19] Asl, B. M. and Mahloojifar, A., "Eigenspace-based minimum variance beamforming applied to medical ultrasound imaging," *IEEE transactions on ultrasonics, ferroelectrics, and frequency control* **57**(11) (2010).

[20] Asl, B. M. and Mahloojifar, A., "Minimum variance beamforming combined with adaptive coherence weighting applied to medical ultrasound imaging," *IEEE transactions on ultrasonics, ferroelectrics, and frequency control* **56**(9) (2009).

[21] Varray, F., Ramalli, A., Cachard, C., Tortoli, P., and Basset, O., "Fundamental and second-harmonic ultrasound field computation of inhomogeneous nonlinear medium with a generalized angular spectrum method," *IEEE transactions on ultrasonics, ferroelectrics, and frequency control* **58**(7) (2011).

[22] Varray, F., Basset, O., Tortoli, P., and Cachard, C., "Creanuis: a non-linear radiofrequency ultrasound image simulator," *Ultrasound in medicine & biology* **39**(10), 1915–1924 (2013).

[23] Ma, Q., Ma, Y., Gong, X., and Zhang, D., "Improvement of tissue harmonic imaging using the pulse-inversion technique," *Ultrasound in medicine & biology* **31**(7), 889–894 (2005).

[24] Holfort, I. K., Gran, F., and Jensen, J. A., "Broadband minimum variance beamforming for ultrasound imaging," *IEEE transactions on ultrasonics, ferroelectrics, and frequency control* **56**(2), 314–325 (2009).

[25] Deylami, A. M. and Asl, B. M., "A fast and robust beamspace adaptive beamformer for medical ultrasound imaging," *IEEE Transactions on Ultrasonics, Ferroelectrics, and Frequency Control* (2017).

[26] Nguyen, M. M., Shin, J., and Yen, J., "Harmonic imaging with fresnel beamforming in the presence of phase aberration," *Ultrasound in medicine & biology* **40**(10), 2488–2498 (2014).

[27] Asl, B. M. and Mahloojifar, A., "A low-complexity adaptive beamformer for ultrasound imaging using structured covariance matrix," *IEEE transactions on ultrasonics, ferroelectrics, and frequency control* **59**(4) (2012).

[28] Moghimirad, E., Hoyos, C. A. V., Mahloojifar, A., Asl, B. M., and Jensen, J. A., "Synthetic aperture ultrasound fourier beamformation using virtual sources," *IEEE transactions on ultrasonics, ferroelectrics, and frequency control* **63**(12), 2018–2030 (2016).

[29] Deylami, A. M. and Asl, B. M., "Low complex subspace minimum variance beamformer for medical ultrasound imaging," *Ultrasonics* **66**, 43–53 (2016).